%% file: bare_conf.tex
\documentclass[conference]{IEEEtran}
\ifCLASSINFOpdf
\else
\fi
\usepackage{multirow}
\usepackage{cite}
\usepackage{amsmath,amssymb,amsfonts}
\usepackage{graphicx}
\usepackage{textcomp}
\usepackage{xcolor}
\usepackage{tabularx}
\usepackage{makecell}
\def\BibTeX{{\rm B\kern-.05em{\sc i\kern-.025em b}\kern-.08em
    T\kern-.1667em\lower.7ex\hbox{E}\kern-.125emX}}

\usepackage{booktabs}
\usepackage{algpseudocode}
\usepackage{algorithm}
\usepackage{amsmath}
\usepackage{amssymb}
\usepackage{multirow}
\usepackage{subcaption}
\usepackage{hyperref}
\usepackage{ulem} 
\usepackage{tikz} 

\usepackage{fancyhdr}
\usepackage{lipsum}  

\fancypagestyle{firstpage}{
  \fancyhf{} 
  \fancyhead[L]{IEEE International Conference on Wearable and Implantable Body Sensor Networks (BSN 2024)} 
  \fancyfoot[C]{\thepage} 
}

\begin{document}
%
\title{MoodPupilar: Predicting Mood Through Smartphone Detected Pupillary Responses in Naturalistic Settings}

\author{
\IEEEauthorblockN{
Rahul Islam \href{https://orcid.org/0000-0003-3601-0078}{\includegraphics[scale=0.06]{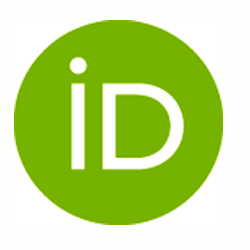}},
Tongze Zhang \href{https://orcid.org/0000-0002-3375-7136}{\includegraphics[scale=0.06]{Fig/orcid.png}},
Priyanshu Singh Bisen \href{https://orcid.org/0009-0007-5433-0020}{\includegraphics[scale=0.06]{Fig/orcid.png}},
Sang Won Bae \href{https://orcid.org/0000-0002-2047-1358}{\includegraphics[scale=0.06]{Fig/orcid.png}}
}
\IEEEauthorblockA{Charles V. Schaefer, Jr. School of Engineering and Science\\
Stevens Institute of Technology\\
Hoboken, USA\\
}
}


%


\maketitle
\thispagestyle{firstpage}

\begin{abstract}
MoodPupilar introduces a novel method for mood evaluation using pupillary response captured by a smartphone's front-facing camera during daily use. Over a four-week period, data was gathered from 25 participants to develop models capable of predicting daily mood averages. Utilizing the GLOBEM behavior modeling platform, we benchmarked the utility of pupillary response as a predictor for mood. Our proposed model demonstrated a Matthew's Correlation Coefficient (MCC) score of 0.15 for Valence and 0.12 for Arousal, which is on par with or exceeds those achieved by existing behavioral modeling algorithms supported by GLOBEM. This capability to accurately predict mood trends underscores the effectiveness of pupillary response data in providing crucial insights for timely mental health interventions and resource allocation. The outcomes are encouraging, demonstrating the potential of real-time and predictive mood analysis to support mental health interventions.
\end{abstract}


%
\IEEEpeerreviewmaketitle

\input{Section/1.Introduction}
\input{Section/2.RelatedWork}
\input{Section/3.Method}
\input{Section/4.Results}

\input{Section/5.Discussion}
\input{Section/7.Conclusion}
\bibliographystyle{IEEEtran}
\bibliography{Reference}

\end{document}

%% file: Section/1.Introduction.tex
\section{Introduction}

Smartphones are central to modern life, enabling everything from communication to managing finances and accessing healthcare services. This ubiquity provides a unique opportunity to monitor mental health by analyzing usage patterns and capturing candid facial expressions and pupillary responses through the front-facing camera. Unlike posed selfies, these candid captures offer a genuine snapshot of emotional states, free from social masking. Advances in machine learning and artificial intelligence enhance affective computing, improving the accuracy of mood detection \cite{likamwa2013moodscope, kang2023k, meegahapola2023generalization, jacob2023affect}. This accuracy facilitates interventions like mood-based music recommendations or prompts for social support, expanding the role of smartphones in human-centered computing, ambient intelligence, and interactive design.

The current state of mood tracking relies on self-reported questions, such as those available on Apple Fitness or Google Fit, which depend on subjective biased recall of one's mood at any given moment. Typically, these reported moods are used as a mood diary for user reference. Most recently, research has been focusing on detecting mood states using sensors that collect personal social and behavioral data. Studies \cite{likamwa2013moodscope, kang2023k, meegahapola2023generalization, jacob2023affect} 
often use data from smartphone app usage, wearable devices, and GPS to predict mood, achieving significant accuracy and F1 scores in classifying various mood states like valence and arousal. This reflects a move towards less intrusive, sensor-based mood detection technologies that offer accurate and immediate evaluations with minimal user effort.

However, mood characterization extends beyond social and behavioral constructs to include physiological signals. Studies indicate that physiological signals from pupillary response correlate with mood states, demonstrating nuanced interactions between pupil dilation and emotional, cognitive processing. For instance, sustained pupil dilation linked to depression shows heightened sensitivity to negative stimuli, decreases after induced negative moods, suggesting a mechanism for maintaining depressive states \cite{steidtmann2010pupil}. These observations in controlled settings highlight potential insights into natural environments.

In our work, we aim to bridge the gap between lab-based research and real-world applications by integrating physiological signals to detect mood using mobile sensing. We employ an affective mobile sensing system \cite{islam2024facepsy, islam2024pupilsense} that operates in the background of a user's smartphone, opportunistically collecting images of the pupil during the user's daily routine. We then use the pupillary response measures computed from these images to create a model for detecting the individual's average daily mood. Our proposed method supports the use of pupillary response as a physiological signal for mood detection, arguing for its incremental utility in mobile sensing when combined with other sensors to enhance mood detection.

%% file: Section/2.RelatedWork.tex
\section{Related Work}
\subsection{Understanding Mood Using Pupillary Response}
In the exploration of mood detection through pupillary response, several recent studies have demonstrated the nuanced relationship between pupil dilation and emotional and cognitive processing. For instance, a study has shown that individuals with a history of depression exhibit more sustained pupil dilation to negative emotional stimuli, a response that tends to decrease following a negative mood induction, indicating a heightened sensitivity and subsequent cognitive blunting in response to prolonged negative stimuli \cite{steidtmann2010pupil}. In social contexts, pupillary contagion—where observers' pupils dilate in response to others' dilated pupils—has been observed regardless of the emotional expression presented, with socioeconomic status influencing the degree of contagion in response to emotionally neutral stimuli \cite{fawcett2022individual}. Moreover, physiological arousal linked to very light exercise has been correlated with pupil dilation, identifying the ascending arousal system's activation at surprisingly low levels of physical exertion \cite{kuwamizu2022pupil}. Furthermore, sustained pupil dilation has been linked to self-reported rumination in depressed individuals, suggesting a potential mechanism for the maintenance of depressive states through prolonged emotional processing \cite{siegle2003seconds}. The most recent advances include employing sophisticated eye-tracking technology to measure pupillary responses during exposure to emotional video scenarios, achieving significant classification accuracies in detecting emotions such as fear, anger, and surprise. These findings underscore the potential of pupillometry in enhancing our understanding of mood and emotional reactivity, promising applications in affective computing and mental health diagnostics. However, this research into pupillary response and mood has primarily taken place in controlled laboratory environments. Nonetheless, observing human behavior and physiological reactions in natural settings can offer distinct insights. 

\subsection{Mood Detection Using Mobile Sensing}
Traditionally, Mood detection via mobile sensing has relied on behavioral and social data gathered from smartphone sensors, with several studies achieving noteworthy accuracies in mood prediction \cite{likamwa2013moodscope, kang2023k, meegahapola2023generalization, jacob2023affect}. For instance, LiKamwa et al. \cite{likamwa2013moodscope} reached up to 67\% accuracy in mood inference using smartphone app usage data to model mood. Kang et al. \cite{kang2023k} achieved an F1 score of 0.543 for valence and 0.534 for arousal utilizing data from both smartphones and wearable devices. More recently, Meegahapola et al. \cite{meegahapola2023generalization} conducted an extensive study to develop a model that classifies valence into high and low categories, achieving an AUC of 0.51. Additionally, Jacob et al. \cite{jacob2023affect} leveraged IMU data from smartphones to predict mood states, obtaining an F1 score of 0.897 for both valence and arousal. These developments highlight a trend towards more unobtrusive, sensor-based mood tracking technologies that reduce the effort required from users while delivering accurate and immediate mood evaluations.

Behavioral data captures long-term trends but often misses rapid emotional changes that physiological signals like pupillary responses can detect. Integrating both data types could improve mood detection systems' robustness. Studies in controlled settings, such as those by Lee et al. \cite{healthcare11030322}, show that eye-tracking metrics can effectively classify emotions with high accuracy. However, the limited real-world applicability of these controlled studies underscores the need for more ecologically valid research. This calls for developing methods that analyze behavioral and physiological data in naturalistic settings to better reflect actual user experiences and behaviors.


%% file: Section/3.Method.tex
\section{Method}
\subsection{Data Collection}
Our research utilized the affective mobile sensing system, FacePsy, which operates in the background on Android smartphones. This system opportunistically collects images of the pupil when the user unlocks their phone or opens any of 35 predefined trigger apps categorized into communication, social media, productivity, entertainment, and health. After activation, FacePsy captures eye images for 10 seconds, processes them to detect and crop the images of the left and right eyes, and transmits these for Pupil-Iris Ratio (PIR) estimation as a measure of pupillary response. This estimation employs the deep learning-based methods detailed in our PupilSense \cite{islam2024pupilsense} research. For further details, please see FacePsy \cite{islam2024facepsy} and PupilSense \cite{islam2024pupilsense}.

\begin{figure*}[h]
  \centering
  \includegraphics[width=17.2cm]{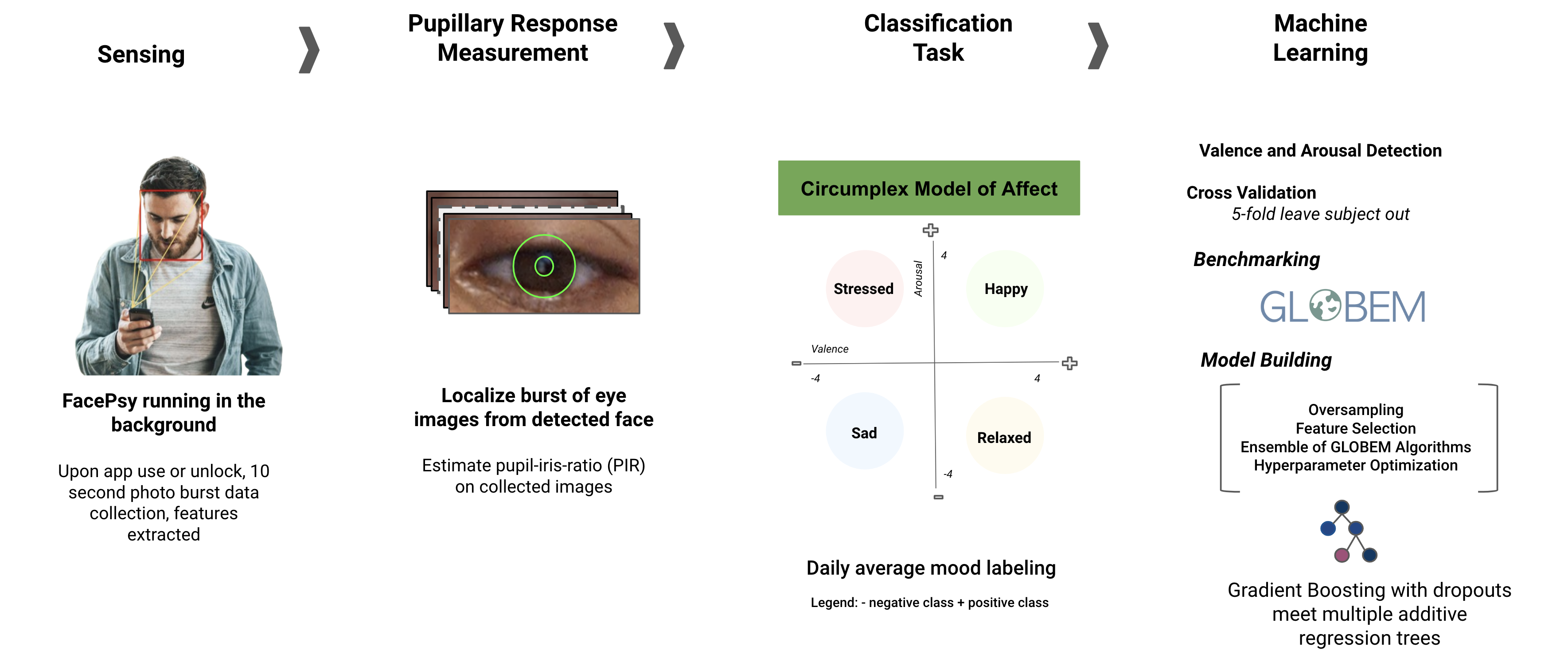}
  \caption{Our mood sensing pipeline through pupillary response.}
  \label{fig:yourlabel}
\end{figure*}

\subsection{Research Design and Participant Demographics}
\subsubsection{Protocol}
The study was conducted remotely during the COVID-19 Public Health Emergency, approved by the Institutional Review Board (IRB). It involved participants aged 18 or older, owning an Android phone. Participants completed baseline activities and engaged with daily mood surveys using the Circumplex Model of Affect (CMA) three times daily for four weeks. Compliance was incentivized with up to \$135 in compensation. Further procedural details are available in the \cite{islam2024facepsy}.

\subsubsection{Participants}
We recruited 38 participants, with 25 completing the study. Reasons for non-completion included excessive battery usage, personal reasons, and failure to meet survey requirements. The participant demographics consisted of 11 males, 8 females, and 6 unspecified, with an average age of 27.88. Most were Asian (15) or Caucasian (4), with varying education levels from high school diplomas to master's degrees. Details on mental health ratings and substance use are reported in the study results.

\subsection{Dataset}
Our system triggered 15,995 data collection instances, capturing eye images during phone interactions. Out of these, 8,299 instances successfully underwent Pupil-Iris Ratio (PIR) estimation, averaging 11.85 daily PIR estimations per participant. We filtered these instances based on the acceptable PIR range of 0.2 (highly constricted) to 0.7 (highly dilated) established by prior research \cite{hollingsworth2009pupil}, resulting in 6,657 usable instances.

Data were segmented into four daily periods—midnight, morning, afternoon, and evening—analyzing statistical features such as sum, minimum, maximum, average, median, and standard deviation of PIR estimations. Data gaps from participant inactivity or errors in estimation were addressed by imputing missing values with daily means. Each participant's daily data included 48 features and was labeled with their average mood scores. The total potential data coverage spanned 616 participant days. Of these, 55 days lacked eye images due to participant-reported unavailability (e.g., vacations), and 33 days were excluded due to poor image quality or non-responsiveness to mood surveys. This left 528 analyzable days in our dataset. Participants completed mood assessments via the CMA survey three times daily over four weeks. Despite collecting 577 days of mood data, due to incomplete daily data, we matched mood reports with PIR data for only 470 days. 

\subsection{Classification Framework}
Our study employs the Circumplex Model of Affect (CMA) to evaluate mood across two primary dimensions: valence and arousal. Each dimension is scored on a scale from -4 to 4. Consistent with methodologies established in prior research such as Likamwa et al. \cite{likamwa2013moodscope} and Meegahapola et al. \cite{meegahapola2023generalization}, we adapt these scales into binary classifications for analytical purposes. Specifically, a score below 0 is categorized as low valence or arousal, whereas a score of 0 or above is categorized as high. This binary classification facilitates the development of mood prediction models that aim to capture the daily averages of valence and arousal. Reliable predictions of daily mood averages could serve multiple purposes. They reflect a user's current emotional state and help identify patterns or shifts in mood, indicating changes from their typical emotional behavior.

\subsubsection{Benchmarking}
With the daily average mood model described above, we benchmark the valence and arousal detection on the GLOBEM Platform \cite{xu2023globem}, which supports the flexible and rapid evaluation of existing behavior modeling methods. We specifically benchmark our dataset using existing machine learning algorithms i.e Canzian et al. \cite{canzian2015trajectories}, Farhan et al. \cite{farhan2016behavior}, Xu et al. \cite{xu2019leveraging}, Lu et al. \cite{lu2018joint}, Saeb et al. \cite{saeb2015mobile}, Wahle et al. \cite{wahle2016mobile}, Wang et al. \cite{wang2018tracking} supported by GLOBEM.

\subsubsection{Our Model}
We further propose our model for valence and arousal detection, which is an ensemble model with feature selection using LightGBM, a gradient boosting algorithm. We employed algorithms supported in GLOBEM \cite{xu2023globem} (i.e Canzian et al. \cite{canzian2015trajectories}, Farhan et al. \cite{farhan2016behavior}, Xu et al. \cite{xu2019leveraging}, Lu et al. \cite{lu2018joint}, Saeb et al. \cite{saeb2015mobile}, Wahle et al. \cite{wahle2016mobile}, Wang et al. \cite{wang2018tracking}) as base learners, using the best hyperparameters. LightGBM serves as the meta learner.

Informed by prior research \cite{xu2023globem} in behavior modeling, we use a \textit{5-fold leave subject out cross validation} to evaluate each model in benchmarking and our proposed model. This ensures that all data from a single participant are exclusively allocated to either the training, validation, or testing phase, but never shared across these subsets. Additionally, we implement hyperparameter optimization in our training dataset for tuning hyperparameters. This subject-independent partitioning approach, combined with cross-validation, enhances the robustness our results compared to methods using a simple train-test split. To evaluate the performance of our model, we use balanced accuracy and Matthew’s Correlation Coefficient (MCC) as our metrics. These metrics were selected because they offer a thorough evaluation of model performance. Specifically, MCC incorporates all elements of the confusion matrix, making it a strong alternative to the F1 score for binary classification tasks, while balanced accuracy equally considers the rates of true positives and true negatives.

%% file: Section/4.Results.tex
\section{Results and Discussion}

Benchmarking our smartphone-collected pupillary response data for mood detection on the GLOBEM Platform against existing behavioral modeling algorithms indicates generally low performance across the board. Notably, the algorithms by Canzian et al. \cite{canzian2015trajectories}, Saeb et al. \cite{saeb2015mobile}, Wahle et al. \cite{wahle2016mobile}, and Wang et al. \cite{wang2018tracking} received negative MCC scores, suggesting that these models might be performing no better than random guessing in some cases, particularly in their ability to reliably predict mood based on the data. In contrast, our model, MoodPupilar, shows improved performance with the highest scores in both Balanced Accuracy and Matthews Correlation Coefficient among the compared models. It achieved a BA of 0.63 and an MCC of 0.15 for valence, and a BA of 0.56 and an MCC of 0.12 for arousal. This indicates a more reliable and effective prediction capability in mood detection using pupillary response compared to the existing approaches on the GLOBEM Platform, which is specifically designed for depression. While individual models on the GLOBEM platform may not perform well, an ensemble model built with the GLOBEM algorithm could offer superior performance.

\begin{table}[h]
\caption{\label{tab:results_cma} Result on the GLOBEM Platform and our model for 5-fold leave-subject-out cross-validation for valence and arousal (BA: Balanced Accuracy, MCC: Matthews correlation coefficients).}
\centering  
\begin{tabular}{p{3cm}p{0.5cm}p{0.9cm}p{0.5cm}p{0.9cm}}
\toprule
\textbf{Model} & \multicolumn{2}{c}{\textbf{Valence}} & \multicolumn{2}{c}{\textbf{Arousal}} \\
\cmidrule(lr){2-3} \cmidrule(lr){4-5}
 & BA & MCC & BA & MCC  \\
\midrule
\multicolumn{5}{c}{\dotuline{GLOBEM Benchmark}} \\
\midrule
Canzian et al. \cite{canzian2015trajectories} & 0.60 & -0.02 & 0.50 &-0.01 \\
Farhan et al. \cite{farhan2016behavior} &0.60 &0.03 &0.53 &0.06 \\
Xu et al. \cite{xu2019leveraging} &0.51 &0.11 &0.55 &0.09 \\
Lu et al. \cite{lu2018joint} &0.63 &0.08 &0.51 &-0.03  \\
Saeb et al. \cite{saeb2015mobile} &0.60 &-0.01 &0.50 &-0.03  \\
Wahle et al. \cite{wahle2016mobile} &0.60 &-0.01 &0.50 &-0.01 \\
Wang et al. \cite{wang2018tracking} &0.60 &-0.01 &0.50 &-0.03 \\
\midrule
\multicolumn{5}{c}{\dotuline{Our Model}} \\
\midrule
MoodPupilar & \textbf{0.63} & \textbf{0.15} & \textbf{0.56} & \textbf{0.12} \\
\bottomrule
\end{tabular}
\end{table}

Current mood-tracking methodologies range from self-reported questionnaires, like those in Apple Fitness or Google Fit, to sensor-based approaches using smartphone and wearable data. Traditional methods rely on users recalling and reporting their mood, often used as a personal mood diary. In contrast, recent research \cite{likamwa2013moodscope, kang2023k, meegahapola2023generalization, jacob2023affect} employs sensors such as smartphone app usage, wearable device metrics, and IMU data to objectively assess mood states, achieving notable accuracies in detecting emotional states such as valence and arousal. This shift towards less intrusive, mobile sensing-based technologies allows for more accurate and immediate mood assessments with reduced user involvement \cite{likamwa2013moodscope, kang2023k, meegahapola2023generalization, jacob2023affect}. However, the current mobile sensing-based solution only explores social and behavioral constructs of mood characterization. Research \cite{fawcett2022individual, steidtmann2010pupil} shows that physiological signals, such as pupillary response, are linked to mood states, revealing complex relationships between pupil dilation and mood. The potential use cases for our findings are extensive. Implementing a model like MoodPupilar could automate mood tracking, providing more accurate, objective, and real-time assessments of mood that reflect a user's immediate emotional state without the need for manual input. This could revolutionize the way mood variations are monitored, offering new insights into mental health and well-being.

%% file: bare_conf.bbl
\begin{thebibliography}{10}
\providecommand{\url}[1]{#1}
\csname url@samestyle\endcsname
\providecommand{\newblock}{\relax}
\providecommand{\bibinfo}[2]{#2}
\providecommand{\BIBentrySTDinterwordspacing}{\spaceskip=0pt\relax}
\providecommand{\BIBentryALTinterwordstretchfactor}{4}
\providecommand{\BIBentryALTinterwordspacing}{\spaceskip=\fontdimen2\font plus
\BIBentryALTinterwordstretchfactor\fontdimen3\font minus \fontdimen4\font\relax}
\providecommand{\BIBforeignlanguage}[2]{{%
\expandafter\ifx\csname l@#1\endcsname\relax
\typeout{** WARNING: IEEEtran.bst: No hyphenation pattern has been}%
\typeout{** loaded for the language `#1'. Using the pattern for}%
\typeout{** the default language instead.}%
\else
\language=\csname l@#1\endcsname
\fi
#2}}
\providecommand{\BIBdecl}{\relax}
\BIBdecl

\bibitem{likamwa2013moodscope}
R.~LiKamWa, Y.~Liu, N.~D. Lane, and L.~Zhong, ``Moodscope: Building a mood sensor from smartphone usage patterns,'' in \emph{Proceeding of the 11th annual international conference on Mobile systems, applications, and services}, 2013, pp. 389--402.

\bibitem{kang2023k}
S.~Kang, W.~Choi, C.~Y. Park, N.~Cha, A.~Kim, A.~H. Khandoker, L.~Hadjileontiadis, H.~Kim, Y.~Jeong, and U.~Lee, ``K-emophone: A mobile and wearable dataset with in-situ emotion, stress, and attention labels,'' \emph{Scientific data}, vol.~10, no.~1, p. 351, 2023.

\bibitem{meegahapola2023generalization}
L.~Meegahapola, W.~Droz, P.~Kun, A.~De~G{\"o}tzen, C.~Nutakki, S.~Diwakar, S.~R. Correa, D.~Song, H.~Xu, M.~Bidoglia \emph{et~al.}, ``Generalization and personalization of mobile sensing-based mood inference models: an analysis of college students in eight countries,'' \emph{Proceedings of the ACM on interactive, mobile, wearable and ubiquitous technologies}, vol.~6, no.~4, pp. 1--32, 2023.

\bibitem{jacob2023affect}
S.~Jacob, P.~Vinod, A.~Subramanian, and V.~G. Menon, ``Affect sensing from smartphones through touch and motion contexts,'' \emph{Multimedia Systems}, vol.~29, no.~5, pp. 2495--2509, 2023.

\bibitem{steidtmann2010pupil}
D.~Steidtmann, R.~E. Ingram, and G.~J. Siegle, ``Pupil response to negative emotional information in individuals at risk for depression,'' \emph{Cognition and Emotion}, vol.~24, no.~3, pp. 480--496, 2010.

\bibitem{islam2024facepsy}
R.~Islam and S.~W. Bae, ``Facepsy: An open-source affective mobile sensing system -- analyzing facial behavior and head gesture for depression detection in naturalistic settings,'' in \emph{Proceedings of the ACM International Conference on Mobile Human-Computer Interaction (MobileHCI)}, 2024.

\bibitem{islam2024pupilsense}
M.~R. Islam and S.~W. Bae, ``Pupilsense: Detection of depressive episodes through pupillary response in the wild,'' in \emph{International Conference on Activity and Behavior Computing}, 2024.

\bibitem{fawcett2022individual}
C.~Fawcett, E.~Nordenswan, S.~Yrttiaho, T.~H{\"a}iki{\"o}, R.~Korja, L.~Karlsson, H.~Karlsson, and E.-L. Kataja, ``Individual differences in pupil dilation to others’ emotional and neutral eyes with varying pupil sizes,'' \emph{Cognition and Emotion}, vol.~36, no.~5, pp. 928--942, 2022.

\bibitem{kuwamizu2022pupil}
R.~Kuwamizu, Y.~Yamazaki, N.~Aoike, G.~Ochi, K.~Suwabe, and H.~Soya, ``Pupil-linked arousal with very light exercise: pattern of pupil dilation during graded exercise,'' \emph{The Journal of Physiological Sciences}, vol.~72, no.~1, p.~23, 2022.

\bibitem{siegle2003seconds}
G.~J. Siegle, S.~R. Steinhauer, C.~S. Carter, W.~Ramel, and M.~E. Thase, ``Do the seconds turn into hours? relationships between sustained pupil dilation in response to emotional information and self-reported rumination,'' \emph{Cognitive Therapy and Research}, vol.~27, pp. 365--382, 2003.

\bibitem{healthcare11030322}
\BIBentryALTinterwordspacing
C.-L. Lee, W.~Pei, Y.-C. Lin, A.~Granmo, and K.-H. Liu, ``Emotion detection based on pupil variation,'' \emph{Healthcare}, vol.~11, no.~3, 2023. [Online]. Available: \url{https://www.mdpi.com/2227-9032/11/3/322}
\BIBentrySTDinterwordspacing

\bibitem{hollingsworth2009pupil}
K.~Hollingsworth, K.~W. Bowyer, and P.~J. Flynn, ``Pupil dilation degrades iris biometric performance,'' \emph{Computer vision and image understanding}, vol. 113, no.~1, pp. 150--157, 2009.

\bibitem{xu2023globem}
X.~Xu, X.~Liu, H.~Zhang, W.~Wang, S.~Nepal, Y.~Sefidgar, W.~Seo, K.~S. Kuehn, J.~F. Huckins, M.~E. Morris \emph{et~al.}, ``Globem: Cross-dataset generalization of longitudinal human behavior modeling,'' \emph{Proceedings of the ACM on Interactive, Mobile, Wearable and Ubiquitous Technologies}, vol.~6, no.~4, pp. 1--34, 2023.

\bibitem{canzian2015trajectories}
L.~Canzian and M.~Musolesi, ``Trajectories of depression: unobtrusive monitoring of depressive states by means of smartphone mobility traces analysis,'' in \emph{Proceedings of the 2015 ACM international joint conference on pervasive and ubiquitous computing}, 2015, pp. 1293--1304.

\bibitem{farhan2016behavior}
A.~A. Farhan, C.~Yue, R.~Morillo, S.~Ware, J.~Lu, J.~Bi, J.~Kamath, A.~Russell, A.~Bamis, and B.~Wang, ``Behavior vs. introspection: refining prediction of clinical depression via smartphone sensing data,'' in \emph{2016 IEEE wireless health (WH)}.\hskip 1em plus 0.5em minus 0.4em\relax IEEE, 2016, pp. 1--8.

\bibitem{xu2019leveraging}
X.~Xu, P.~Chikersal, A.~Doryab, D.~K. Villalba, J.~M. Dutcher, M.~J. Tumminia, T.~Althoff, S.~Cohen, K.~G. Creswell, J.~D. Creswell \emph{et~al.}, ``Leveraging routine behavior and contextually-filtered features for depression detection among college students,'' \emph{Proceedings of the ACM on Interactive, Mobile, Wearable and Ubiquitous Technologies}, vol.~3, no.~3, pp. 1--33, 2019.

\bibitem{lu2018joint}
J.~Lu, C.~Shang, C.~Yue, R.~Morillo, S.~Ware, J.~Kamath, A.~Bamis, A.~Russell, B.~Wang, and J.~Bi, ``Joint modeling of heterogeneous sensing data for depression assessment via multi-task learning,'' \emph{Proceedings of the ACM on Interactive, Mobile, Wearable and Ubiquitous Technologies}, vol.~2, no.~1, pp. 1--21, 2018.

\bibitem{saeb2015mobile}
S.~Saeb, M.~Zhang, C.~J. Karr, S.~M. Schueller, M.~E. Corden, K.~P. Kording, D.~C. Mohr \emph{et~al.}, ``Mobile phone sensor correlates of depressive symptom severity in daily-life behavior: an exploratory study,'' \emph{Journal of medical Internet research}, vol.~17, no.~7, p. e4273, 2015.

\bibitem{wahle2016mobile}
F.~Wahle, T.~Kowatsch, E.~Fleisch, M.~Rufer, S.~Weidt \emph{et~al.}, ``Mobile sensing and support for people with depression: a pilot trial in the wild,'' \emph{JMIR mHealth and uHealth}, vol.~4, no.~3, p. e5960, 2016.

\bibitem{wang2018tracking}
R.~Wang, W.~Wang, A.~DaSilva, J.~F. Huckins, W.~M. Kelley, T.~F. Heatherton, and A.~T. Campbell, ``Tracking depression dynamics in college students using mobile phone and wearable sensing,'' \emph{Proceedings of the ACM on Interactive, Mobile, Wearable and Ubiquitous Technologies}, vol.~2, no.~1, pp. 1--26, 2018.

\end{thebibliography}
